\documentclass{PoS}

\title{Search for the Charmonium Dissociation Temperature with Variational Analysis in Lattice QCD }

\ShortTitle{Charmonium Dissociation Temperature with Variational Analysis}

\author{\speaker{Hiroshi Ohno},  Takashi Umeda, and Kazuyuki Kanaya (WHOT-QCD Collaboration)\\
       Graduate School of Pure and Applied Sciences, University of Tsukuba,
       Tsukuba, Ibaraki 305-8571, Japan \\
       E-mail: \email{ohno@het.ph.tsukuba.ac.jp}}

\abstract{
Charmonium dissociation temperatures are studied in a
quenched anisotropic lattice QCD with standard plaquette
gauge action and $O(a)$ improved Wilson fermion action.
Simulations are carried out at temperatures in the range $0.88\;T_c$
to $2.3\;T_c$. 
From the meson correlators, we first subtract the contribution of constant mode, 
which was reported to mislead the analysis of, in particular, P wave signals.
We then calculate effective masses and Bethe-Salpeter wave functions for 
ground (1S, 1P) and excited states (2S, 2P) using the multi-state variational analysis. 
To distinguish between bound states and scattering states, we
apply two methods: First, we compare effective masses for charmonium
correlation functions with finite spatial momenta
under different spatial boundary conditions.
Since the scattering state energies are sensitive to the boundary
conditions, we expect finite volume effects when the charmonium
dissociates. Second, we study the Bethe-Salpeter wave function, which
should become broad when the charmonium bound state turns into a scattering
state.
With both methods, we have fond no clear evidences of dissociation for the ground and exited 
charmonium states up to $2.3\;T_c$ so far.
}

\FullConference{The XXVI International Symposium on Lattice Field Theory \\
                July 14 - 19, 2008\\
                Williamsburg, Virginia, USA}

\begin{document}

\section{Introduction}
Studying heavy quarkonia properties above $T_c$ is important to
understand how quark-gluon plasma (QGP) is formed 
in heavy ion collisions. In particular, dissociation of charmonia in
deconfined phase is an interesting subject  
because $J/\psi$ suppression \cite{Matsui_Satz} is one of the most
important signals of QGP.
Recently, the sequential $J/\psi$ suppression scenario \cite{seq_Jpsi1}, 
in which dissociation of heavier charmonia ($\chi_c$, $\psi^\prime$) 
play important roles for the suppression of total $J/\psi$ yield, 
has been proposed to understand the experimental results in heavy ion
collisions at SPS and RHIC.

Lattice QCD studies of charmanium spectral function using the
Maximum Entropy Method (MEM) suggested that S wave charmonia, such as
$\eta_c$ and $J/\psi$, survive up to $1.5\;T_c$ \cite{lqcd1,lqcd2,lqcd3,lqcd4,lqcd5,lqcd6}, 
while P wave charmonia, such as $\chi_c$'s, dissociate just above $T_c$ \cite{lqcd4,lqcd5,lqcd6}. 

Although these results look supporting the sequential 
$J/\psi$ suppression scenario, there are still some problems.
First, the meson correlators receive sizable
contribution of the constant mode above $T_c$, especially in P wave correlators\cite{const_mode}.
From an analysis in which the constant mode contribution was properly taken into account, the
thermal effects in charmonium correlators including those of P waves turned out to be quite small up to 1.4 $T_c$.
This is in contrast to the conclusion of previous MEM studies on P wave charmonia. 
Furthermore, choice of the default model for the spectral function is known to affect the MEM results rather sensitively.
A crosscheck without  Bayesian-type analyses is needed.

Let us first discuss what we expect on the lattice when a charmonium dissociates. 
It should be noted that, on a lattice with finite extent, spectral functions consist of discrete spectra only, also at $T>T_c$.
(These discrete spectra appear as broad peaks in approximate calculations such as MEM.)
Below $T_c$, we expect to have discrete peaks corresponding to ground and exited bound states.
When a charmonium bound state fully dissolves above $T_c$, 
the peak corresponding to the bound state will 
vanish and other peaks corresponding $c$-$\bar{c}$ scattering states 
may appear near the vanished peak.
Furthermore, we expect that the wave function for the peak will have a
localized shape when the peak corresponds to a bound state, while it
will have a broad shape extending to large distances when the peak
corresponds to a scattering state.

In this paper, we report on our study of charmonium dissociation.
In principle, when a high precision data of the correlation function is available up to large distances, the constant mode can be identified by conventional analyses too.
In practice, however, with current accuracy and the range of sensible data, we think that it is safer to explicitly subtract out the contribution of the constant mode from the correlators.
We apply the midpoint subtraction method developed in \cite{const_mode}.
Avoiding Bayesian-type analyses, 
we then adopt the multi-state variational method to extract ground and exited state masses both in S and P waves \cite{var}.
At the same time, we calculate the Bethe-Salpeter wave functions of these states  from the spatial correlation function between $c$-$\bar{c}$ quarks \cite{lqcd1}. 

To distinguish bound states of $c$-$\bar{c}$ quarks from their
scattering states for the extracted states, we study the spatial boundary condition dependence of the
energy spectrum at finite volume\cite{Iida}:
The energy of scattering states depends on its relative momentum 
which is quantized according to the spatial size and boundary conditions.
On the other hand the spectrum of the bound states does not change
against such exchange of boundary conditions.
The Bethe-Salpeter wave functions for a bound state will be compact, while
those for scattering state will extend and will change its shape
depending on the spatial lattice size. 
Combining these tests, we may examine if the charmonium is dissociated.

\section{Multi-state variational analysis}

The charmonia correlation matrix in Euclidian space-time is defined by 
\begin{equation} 
C_{ij}(t)\equiv \sum_{\vec{x}}\langle
O^{\Gamma}_i(\vec{x},t)O^{\Gamma}_j(\vec{0},0)^{\dag} \rangle . 
\label{Cij}
\end{equation}
with $\Gamma=\gamma_5$, $\gamma_i$, $\mathbf{1}$
($i=1,\;2,\;3$) for pseudo-scalar (Ps), vector (Ve) and scalar (Sc) channels, respectively. 
Here $\{O^{\Gamma}_i |\, i=1,\;2,\;\cdots N_{\mathrm{state}}\}$ are
smeared meson operators defined by 
\begin{equation}
O^{\Gamma}_i(\vec{x},t)\equiv
\sum_{\vec{y},\vec{z}}\omega_i(\vec{y})\omega_i(\vec{z})\bar{q}(\vec{x}+\vec{y},t)\Gamma
q(\vec{x}+\vec{z},t), 
\end{equation}
with Gaussian smearing functions $\omega_i(\vec{x})\equiv
\exp(-A_i\vert\vec{x}\vert^2)$. 
Table \ref{t1} shows our choice of the smearing parameters $A_i$.

\begin{table}
\caption{The parameter $A_i$ of the smearing functions.}
\label{t1}
\begin{center}
\begin{tabular}{|c||c|c|c|c|c|c|} \hline
 $i$  &   1  &   2  &   3  &   4  &   5  &   6  \\ \hline
$A_i$ & 0.02 & 0.05 & 0.10 & 0.15 & 0.20 & 0.25 \\ \hline
\end{tabular}
\end{center}
\end{table}

According to \cite{const_mode}, constant mode effects are large 
in the deconfined phase, 
because the constant mode is due to wraparound contributions of single quark
propagators, which are suppressed in the confined phase. 
To separate out the constant mode contribution from meson correlators, we study 
midpoint subtracted correlators $\overline{C}(t)$ defined by
\begin{eqnarray}
\overline{C}(t) &\equiv& C(t)-C(N_t/2) \nonumber\\
 &=&
 \left(c_0+\sum^{\infty}_{k=1}c_k\;\cosh\left[m_k\left(t-\frac{N_t}{2}\right)\right]\right)-\sum^{\infty}_{k=0}c_k \nonumber \\  
 &=& 2\sum^{\infty}_{k=1}
 c_k\;\sinh^2\left[\frac{m_k}{2}\left(t-\frac{N_t}{2}\right)\right]. 
\end{eqnarray}
Here $C(t)$ is the matrix of $C_{ij}(t)$ defined in (\ref{Cij}), $c_0$ is a contribution of the constant mode and $N_t$ is the temporal lattice size.  

We extract ground and excited states in $\overline{C}$ by solving the eigenvalue problem,
\begin{equation}
\overline{C}(t)\vec{v}_k=\lambda_k(t;t_0)\overline{C}(t_0)\vec{v}_k
\qquad (k=1,\;2,\;\cdots,\,N_{\mathrm{state}}).
\end{equation}
Effective masses of the diagonalized states are given by
\begin{equation}
\lambda_k(t;t_0)=\frac{\sinh^2\left[\frac{M^{\mathrm{eff}}_k}{2}
\left(t-\frac{N_t}{2}\right)\right]}{\sinh^2\left[\frac{M^{\mathrm{eff}}_k}{2}
\left(t_0-\frac{N_t}{2}\right)\right]}.  
\end{equation}

Next, we define wave functions in terms of the Bethe-Salpeter (BS) amplitude:
\begin{eqnarray}
BS_i(\vec{r},t) &\equiv& \sum_{\vec{x}}\langle
\bar{q}(\vec{x},t)\Gamma_{\mathrm{snk}}q(\vec{x}+\vec{r},t)
O^{\Gamma_{\mathrm{src}}}_i(\vec{0},0)^\dag
\rangle \nonumber \\ 
 &=& \sum^{\infty}_{k=0}
 (\psi_k(\vec{r}))_i\cosh\left[m_k\left(t-\frac{N_t}{2}\right)\right]. 
\end{eqnarray}
Here $(\psi_k(\vec{r}))_i$ is the wave function for the $k$-th state created by the smeared source operator $O^{\Gamma_{\mathrm{src}}}_i$.
To study S waves, we adopt 
$\Gamma_{\mathrm{src}}=\Gamma_{\mathrm{snk}}=\gamma_5$ and $\gamma_i$ for
Ps and Ve charmonia, respectively. 
For Sc charmonium in P wave, we use a
derivative source $\Gamma_{\mathrm{src}}= \gamma_i
(\overleftarrow{\partial_i}-\overrightarrow{\partial_i})$ to enhance the signal, 
with 
$\Gamma_{\mathrm{snk}}= \gamma_i $.

To seprate out the constant mode, we define midpoint subtracted BS amplitudes by
\begin{equation}
\overline{BS}(\vec{r},t)\equiv BS(\vec{r},t)-BS(\vec{r},N_t/2).
\end{equation}
Finally, the diagonalized wave functions are given by
\begin{equation}
\Psi_k(\vec{r};\vec{r_0})\equiv \frac{\sum_i
\overline{BS}_i(\vec{r},t) V_{ik}}{\sum_i
\overline{BS}_i(\vec{r_0},t) V_{ik}}. 
\label{eq:Psi}
\end{equation}
where $V_{ik}\equiv (v_k)_i$ is the $i$-th component of the eigen vector $\vec{v}_k$, 
and $\vec{r_0}$ is the normalization point.

\section{Numerical results}
\subsection{Lattice setup}
We study quenched QCD with the anisotropy $\xi\equiv a_s/a_t=4$. 
The gauge action is the standard plaquette action with $\beta=6.10$
and a bare anisotropy parameter $\gamma_G=3.2108$. 
For quarks, we adopt $O(a)$ improved Wilson fermion action with
tadpole-improved tree-level clover coefficients. 
Our actions and their parameters are the same as those in \cite{action} except for the choice of the Wilson parameter, $r=1$. 
The spatial lattice spacing is $a^{-1}_s=2.030(13)$ GeV. 
Simulations are performed on $N^3_s*N_t$ lattice, with
$N_s=16,\;20$ and $32$ and $N_t=32$, 26, 20, 16 and 12, which
correspond to 
$T = (N_t a_t)^{-1} = 0.88\;T_c$, $1.1\;T_c$, $1.4\;T_c$, $1.8\;T_c$ and $2.3\;T_c$
respectively. (Critical temperature $T_c$ corresponds to $N_t\simeq 28$.) 
The number of gauge configurations is shown in
Table \ref{t2}. 
We choose the Coulomb gauge fixing to calculate the wave functions.  

\begin{table}
\caption{The number of gauge configurations.}
\label{t2}
\begin{center}
\begin{tabular}{|c||c|c|c|c|c|c|} \hline
                     & $N_s=16$ & $N_s=20$ & $N_s=32$ \\ \hline\hline
 local operator      &   $800$  &   $800$  &   $200$  \\ \hline
 derivative operator &   $300$  &   $300$  &   $200$  \\ \hline
\end{tabular}
\end{center}
\end{table}

\subsection{Effective masses}

\begin{figure}[tbp]
 \begin{center}
  \includegraphics[width=80mm, angle=-90]{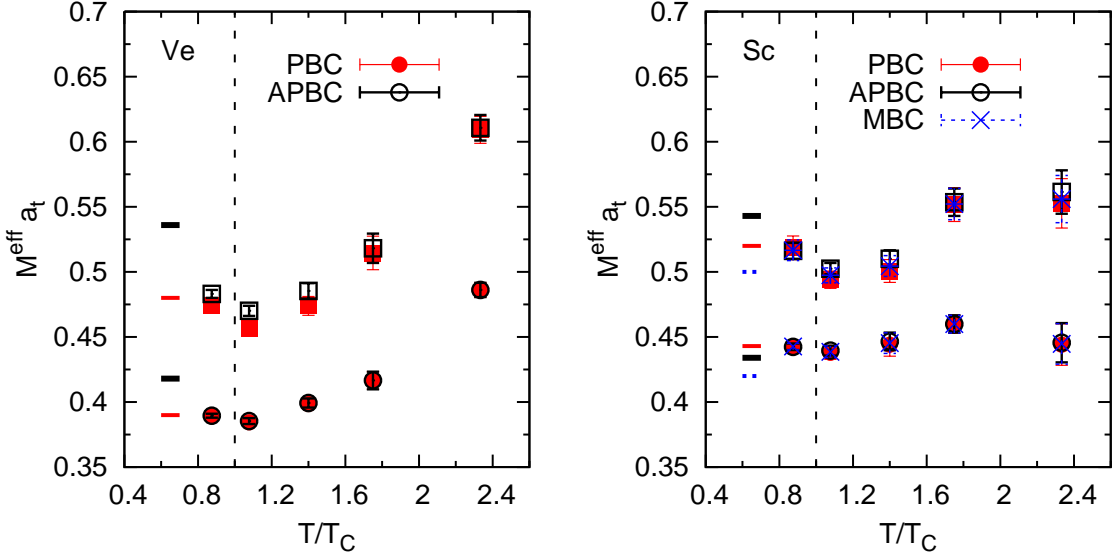}
  \caption{The effective masses for Ve (left) and Sc (right) charmonia. 
  Lower and upper symbols correspond to the ground and first excited states, respectively.
 Red, black and blue symbols indicate the effective masses with PBC, APBC and MBC,
  respectively (see text for the definition of MBC). 
  Short bars shown near the left end of each figure indicate the boundary condition dependence of energy spectra for the case of free quarks. 
} 
  \label{f1}
 \end{center}
\end{figure} 

In Figure \ref{f1} we compare the ground state's and the first excited state's
effective masses of Ve and Sc charmonia as functions of $T$ between different boundary conditions, i.e.\
periodic (PBC), anti-periodic (APBC) and a mixed (MBC) boundary conditions. 
Here MBC is defined by APBC in the $x$-direction combined with PBC in $y$ and $z$-directions.
At zero temperature, the ground state and the first
excited state of Ve channel correspond to
$J/\psi$ and $\psi^\prime$ respectively, and the ground state of Sc
channel corresponds to $\chi_{c0}$. 
The effective masses are calculated by the multi-state variational analysis 
with $4\times 4$ correlation matrix on the $20^3 \times N_t$ lattice. 
Here we choose smeared operators with the parameter $A_3,\;A_4,\;A_5$
and $A_6$ of Table \ref{t1} to obtain the largest overlap with the ground state and the
first excited state. 

To see a typical magnitude of the boundary-condition dependence in the effective masses, 
we have studied the energy spectra for the case of free quark in a box of $(2\;\mathrm{fm})^3$, 
assuming that the charm quark mass is $1.3$ GeV.
Results are shown by short bars around the left end of each plot in Fig.\ref{f1}.
(Only relative differences in the vertical coordinates are relevant
for the bars.)
For ground state P waves, we obtain the largest gap between PBC and MBC. 
Therefore, we may expect a mass shift of about 200 MeV (500 MeV) for a ground (first exited) state when the charmonium fully dissolve.

In Fig.\ref{f1}, we find no evidence of mass shift with such magnitudes, for neither the
ground states nor the first excited states, up to a quite high temperature of about 2.3 $T_c$. 
The results for Ps charmonia are similar.
This suggests that these charmonia bound state persist up to about 2.3 $T_c$. 

\subsection{Wave functions}

\begin{figure}[tbp]
 \begin{center}
  \includegraphics[width=80mm, angle=-90]{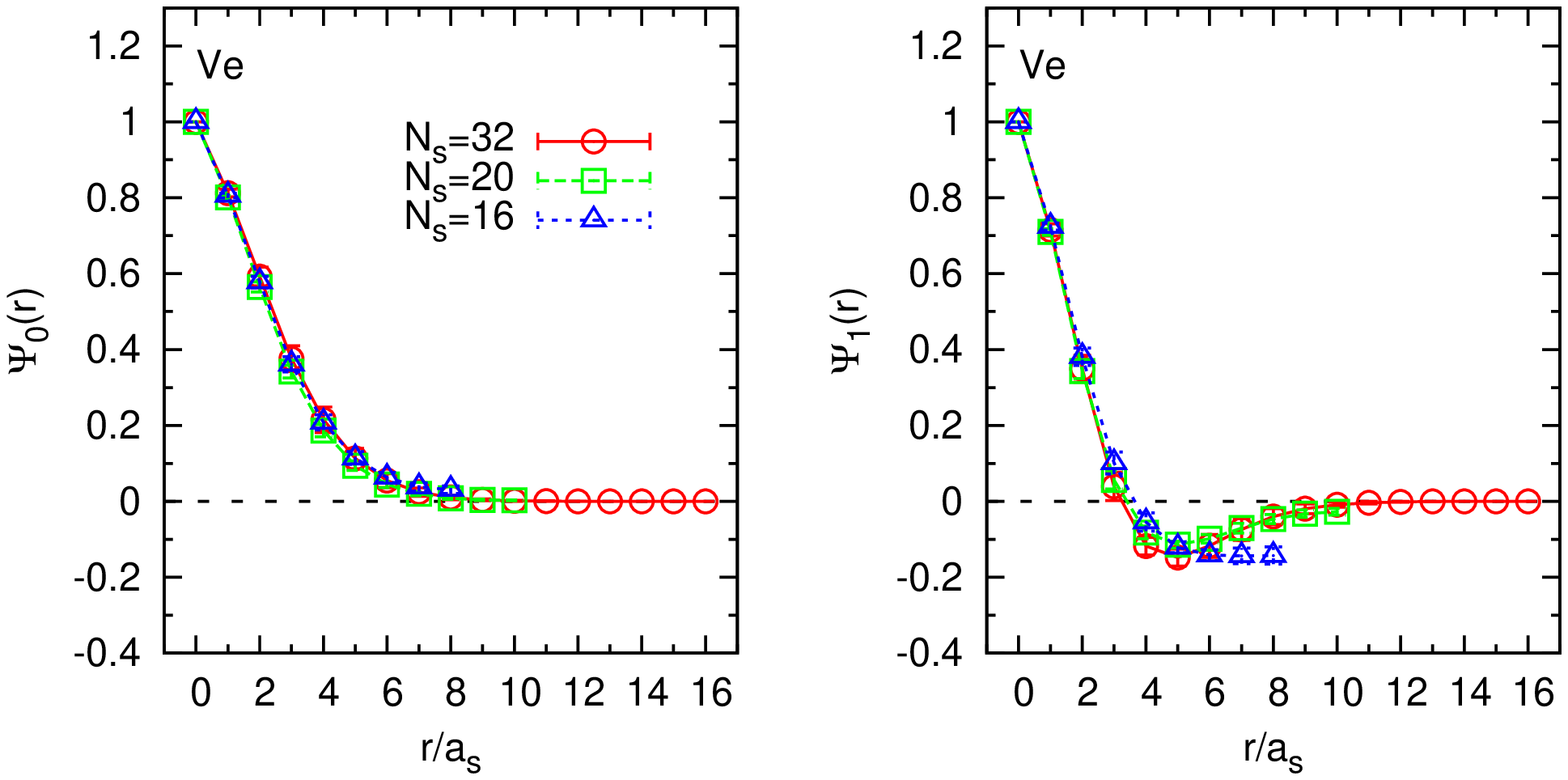}
  \includegraphics[width=80mm, angle=-90]{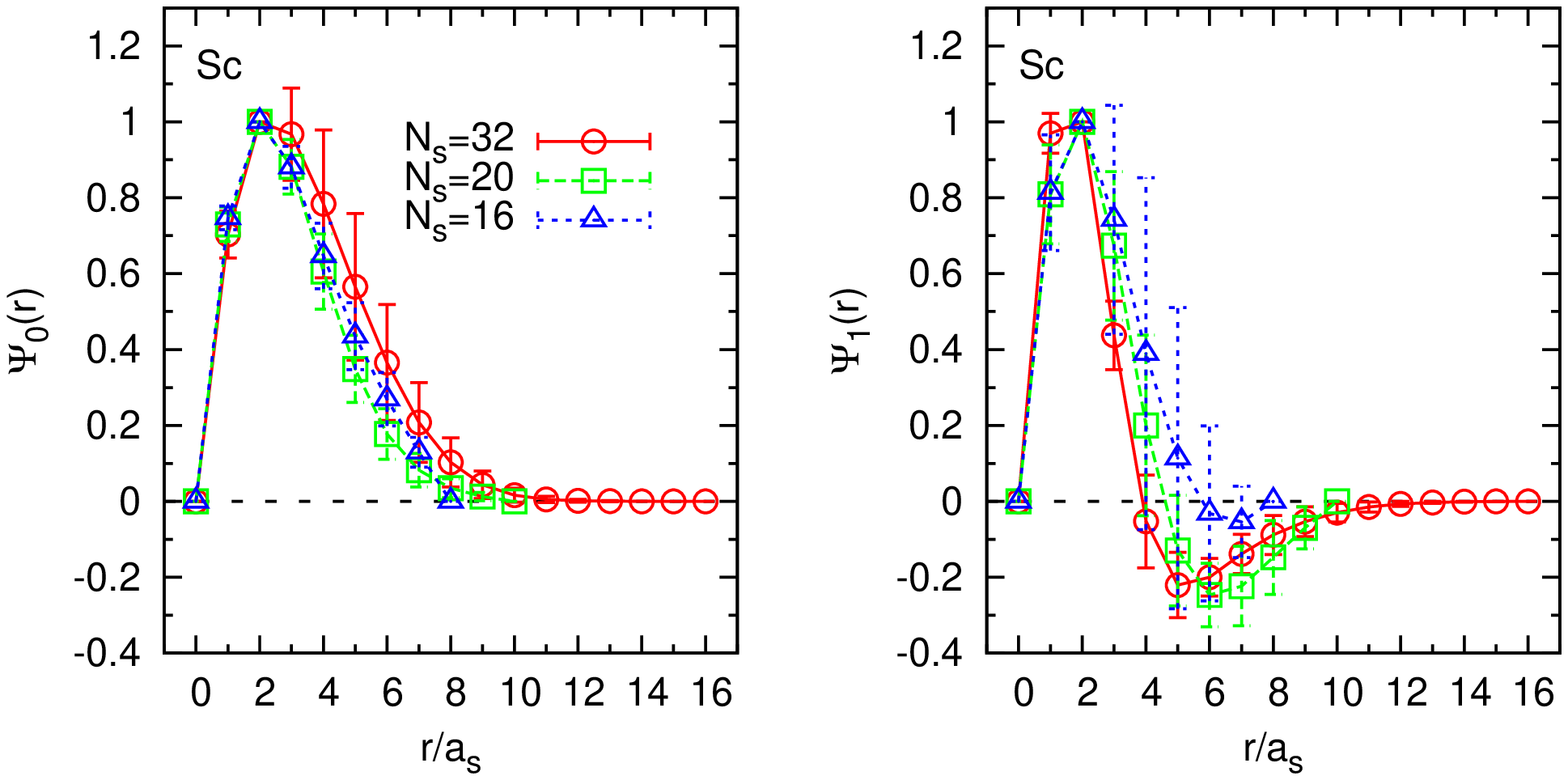}
  \caption{The wave functions for Ve (top) and Sc (bottom) charmonia at 2.3 $T_C$. 
  Horizontal axis is the spatial distance in the $x$-direction.
  Left panels show the results for ground states
  and right panels show those of the first excited states. 
 Red, green and blue symbols indicate data obtained on $N_s=32$, 20 and 16
  lattices.} 
  \label{f2}
 \end{center}
\end{figure} 

To check the bound-state characteristics of the charmonium states, we study the wave functions corresponding to the mass eigenstates diagonalized by the multi-state variational analysis discussed in the previous subsection. 
When a charmonium dissociates, corresponding wave function should extend to large distances and show a sensitive dependence on the spatial size of the box.
Figure \ref{f2} shows the ground state's and the first excited state's
wave functions defined by (\ref{eq:Psi}) for Ve and Sc charmonia at 2.3 $T_c$, obtained on three different spatial lattices of $32^3$, $20^3$ and $16^3$.
We find that, for all charmonia we study, the spatial size dependence is small, and the wave functions are compact. 
We thus find no sign of $c$-$\bar{c}$ scattering states up to 2.3 $T_c$, in accordance with the observation discussed in the previous subsection.

\section{Conclusions}
We investigated if charmonia dissociates in high temperature QCD on quenched anisotropic lattices.  
Adopting the multi-state variational analysis to extract ground and first excited states, we examined spatial boundary condition dependence of
effective masses and investigated the shape and spatial size dependence of wave functions for these states.  
From these studies, we found no sign of scattering states nor clear evidence of 
dissociation, 
both for the ground and first exited states of charmonia
in Ps, Ve and Sc channels up to a quite high temperature of 2.3 $T_c$, so far.

\vspace{5mm}  
We thank the members of the WHOT-QCD Collaboration for innumerable discussions and suggestions. 
This work is in part supported by Grants-in-Aid of the Japanese Ministry
of Education, Culture, Sports, Science and Technology
(Nos.~17340066 and 19549001). 
Numerical calculations were performed on supercomputers at RCNP, Osaka University.

\end{document}